\documentclass[amsmath,amssymb,twocolumn,prc,showpacs,nofootinbib,floatfix,superscriptaddress]{revtex4-1}
\usepackage{graphics}
\usepackage{dcolumn}
\usepackage {bm}
\usepackage[pdftex]{graphicx}
\usepackage{xcolor}
\usepackage{dcolumn}
\usepackage{footnote}
\usepackage{amssymb}
\usepackage{multirow}
\usepackage{natbib}
\usepackage{xcolor}

\newcommand{\A}[1]{$^{#1}$} %% a command to give mass numbers for nuclei
\newcommand{\pg}{($p,\gamma$)} %% a command to write (p,y) instead of making me type it every time

\begin{document}

\title{Isobaric multiplet mass equation in the $A=31$ $T = 3/2$ quartets}
\author{M.~B.~Bennett}
\email{bennettm@nscl.msu.edu}
\affiliation{Department of Physics and Astronomy, Michigan State University, East Lansing, Michigan 48824, USA}
\affiliation{National Superconducting Cyclotron Laboratory, Michigan State University, East Lansing, Michigan 48824, USA}
\affiliation{Joint Institute for Nuclear Astrophysics, Michigan State University, East Lansing, Michigan 48824, USA}
\author{C.~Wrede}
\email{wrede@nscl.msu.edu}
\affiliation{Department of Physics and Astronomy, Michigan State University, East Lansing, Michigan 48824, USA}
\affiliation{National Superconducting Cyclotron Laboratory, Michigan State University, East Lansing, Michigan 48824, USA}
\author{B.~A.~Brown}
\affiliation{Department of Physics and Astronomy, Michigan State University, East Lansing, Michigan 48824, USA}
\affiliation{National Superconducting Cyclotron Laboratory, Michigan State University, East Lansing, Michigan 48824, USA}
\author{S.~N.~Liddick}
\affiliation{National Superconducting Cyclotron Laboratory, Michigan State University, East Lansing, Michigan 48824, USA}
\affiliation{Department of Chemistry, Michigan State University, East Lansing, Michigan 48824, USA}
\author{D.~P\'erez-Loureiro}
\affiliation{Department of Physics and Astronomy, Michigan State University, East Lansing, Michigan 48824, USA}
\affiliation{National Superconducting Cyclotron Laboratory, Michigan State University, East Lansing, Michigan 48824, USA}
\author{D.~W.~Bardayan}
\affiliation{Department of Physics, University of Notre Dame, Notre Dame, Indiana 46556, USA}
\author{A.~A.~Chen}
\affiliation{Department of Physics and Astronomy, McMaster University, Hamilton, Ontario L8S 4M1, Canada}
\author{K.~A.~Chipps}
\affiliation{Oak Ridge National Laboratory, Oak Ridge, Tennessee 37831, USA}
\affiliation{Department of Physics and Astronomy, University of Tennessee, Knoxville, Tennessee 37996, USA}
\author{C.~Fry}
\affiliation{Department of Physics and Astronomy, Michigan State University, East Lansing, Michigan 48824, USA}
\affiliation{National Superconducting Cyclotron Laboratory, Michigan State University, East Lansing, Michigan 48824, USA}
\affiliation{Joint Institute for Nuclear Astrophysics, Michigan State University, East Lansing, Michigan 48824, USA}
\author{B.~E.~Glassman}
\affiliation{Department of Physics and Astronomy, Michigan State University, East Lansing, Michigan 48824, USA}
\affiliation{National Superconducting Cyclotron Laboratory, Michigan State University, East Lansing, Michigan 48824, USA}
\author{C.~Langer}
\affiliation{National Superconducting Cyclotron Laboratory, Michigan State University, East Lansing, Michigan 48824, USA}
\affiliation{Joint Institute for Nuclear Astrophysics, Michigan State University, East Lansing, Michigan 48824, USA}
\author{N.~R.~Larson}
\affiliation{National Superconducting Cyclotron Laboratory, Michigan State University, East Lansing, Michigan 48824, USA}
\affiliation{Department of Chemistry, Michigan State University, East Lansing, Michigan 48824, USA}
\author{E.~I.~McNeice}
\affiliation{Department of Physics and Astronomy, McMaster University, Hamilton, Ontario L8S 4M1, Canada}
\author{Z.~Meisel}
\affiliation{Joint Institute for Nuclear Astrophysics, Michigan State University, East Lansing, Michigan 48824, USA}
\affiliation{Department of Physics, University of Notre Dame, Notre Dame, Indiana 46556, USA}
\author{W.~Ong}
\affiliation{Department of Physics and Astronomy, Michigan State University, East Lansing, Michigan 48824, USA}
\affiliation{National Superconducting Cyclotron Laboratory, Michigan State University, East Lansing, Michigan 48824, USA}
\affiliation{Joint Institute for Nuclear Astrophysics, Michigan State University, East Lansing, Michigan 48824, USA}
\author{P.~D.~O'Malley}
\affiliation{Department of Physics, University of Notre Dame, Notre Dame, Indiana 46556, USA}
\author{S.~D.~Pain}
\affiliation{Oak Ridge National Laboratory, Oak Ridge, Tennessee 37831, USA}
\author{C.~J.~Prokop}
\affiliation{National Superconducting Cyclotron Laboratory, Michigan State University, East Lansing, Michigan 48824, USA}
\affiliation{Department of Chemistry, Michigan State University, East Lansing, Michigan 48824, USA}
\author{S.~B.~Schwartz}
\affiliation{Department of Physics and Astronomy, Michigan State University, East Lansing, Michigan 48824, USA}
\affiliation{National Superconducting Cyclotron Laboratory, Michigan State University, East Lansing, Michigan 48824, USA}
\affiliation{Department of Geology and Physics, University of Southern Indiana, Evansville Indiana 47712, USA}
\author{S.~Suchyta}
\affiliation{National Superconducting Cyclotron Laboratory, Michigan State University, East Lansing, Michigan 48824, USA}
\affiliation{Department of Chemistry, Michigan State University, East Lansing, Michigan 48824, USA}
\author{P.~Thompson}
\affiliation{Department of Physics and Astronomy, University of Tennessee, Knoxville, Tennessee 37996, USA}
\author{M.~Walters}
\affiliation{Department of Physics and Astronomy, McMaster University, Hamilton, Ontario L8S 4M1, Canada}
\author{X.~Xu}
\affiliation{Department of Physics and Astronomy, Michigan State University, East Lansing, Michigan 48824, USA}
\affiliation{National Superconducting Cyclotron Laboratory, Michigan State University, East Lansing, Michigan 48824, USA}
%
%%\date{\today}

\begin{abstract}

\textbf{Background:} The observed mass excesses of analog nuclear states with the same mass number $A$ and isospin $T$ can be used to test the isobaric multiplet mass equation (IMME), which has, in most cases, been validated to a high degree of precision.  A recent measurement [Kankainen \textit{et al.}, Phys. Rev. C \textbf{93} 041304(R) (2016)] of the ground-state mass of \A{31}Cl led to a substantial breakdown of the IMME for the lowest $A = 31, T = 3/2$ quartet. The second-lowest $A = 31, T = 3/2$ quartet is not complete, due to uncertainties associated with the identity of the \A{31}S member state.  

\textbf{Purpose:} Our goal is to populate the two lowest $T = 3/2$ states in \A{31}S and use the data to investigate the influence of isospin mixing on tests of the IMME in the two lowest $A = 31, T = 3/2$ quartets.

\textbf{Methods:} Using a fast \A{31}Cl beam implanted into a plastic scintillator and a high-purity Ge $\gamma$-ray detection array, $\gamma$ rays from the \A{31}Cl$(\beta\gamma)$\A{31}S sequence were measured.  Shell-model calculations using USDB and the recently-developed USDE interactions were performed for comparison.

\textbf{Results:}  Isospin mixing between the \A{31}S isobaric analog state (IAS) at 6279.0(6) keV and a nearby state at 6390.2(7) keV was observed.  The second $T = 3/2$ state in \A{31}S was observed at $E_x = 7050.0(8)$ keV.  Calculations using both USDB and USDE predict a triplet of isospin-mixed states, including the lowest $T = 3/2$ state in \A{31}P, mirroring the observed mixing in \A{31}S, and two isospin-mixed triplets including the second-lowest $T = 3/2$ states in both \A{31}S and \A{31}P.

\textbf{Conclusions:} Isospin mixing in \A{31}S does not by itself explain the IMME breakdown in the lowest quartet, but it likely points to similar isospin mixing in the mirror nucleus \A{31}P, which would result in a perturbation of the \A{31}P IAS energy.  USDB and USDE calculations both predict candidate \A{31}P states responsible for the mixing in the energy region slightly above $E_x = 6400$ keV.  The second quartet has been completed thanks to the identification of the second \A{31}S $T = 3/2$ state, and the IMME is validated in this quartet.

\vskip\baselineskip

\noindent \hfill \scriptsize{PACS numbers: 21.10.Hw, 21.60.Cs, 23.20.Lv, 27.30.+t}

\end{abstract}

\maketitle

\section{Introduction}
\label{intro}
Due to the charge-independent nature of the strong nuclear force, it is possible to model the proton and neutron as spin-like ``isospin'' states of a single particle, the nucleon.  This isospin model treats both the proton and the neutron as degenerate particles with isospin $T = 1/2$, but with opposite isospin projections: $T_z = +1/2$ for neutrons and $T_z = -1/2$ for protons \cite{HeisenbergZPhys1932}.  Thus, nuclei that share a given total mass number $A$ can be seen as total projection states, each with $T_z = (N - Z) / 2$, where $N$ and $Z$ are the number of neutrons and protons, respectively.  Each energy level in a given nucleus itself possesses a total isospin $T$, so it is possible to treat analogous states in isobaric nuclei as members of a $(2T+1)$-member multiplet, each with the same $T$ and a different isospin projection $T_z$.

Under this symmetric formalism, analogous energy states  with the same isospin have exactly the same mass excess values $\Delta$.  However, electrostatic effects perturb the energies of nuclear analog states with differing numbers of protons, breaking this degeneracy and resulting in systematically different energies for multiplet members.  First proposed by Wigner, the isobaric multiplet mass equation (IMME) \cite{WignerProceedings1957,WeinbergPR1959} is a model that uses first-order perturbation theory to predict that the mass excesses of nuclear isobaric analog states (IAS) within an isospin multiplet are systematically related by their isospin projections $T_z$ according to the following quadratic equation:

\begin{equation}
\Delta(T_z) = a + bT_z + cT_z^2
\end{equation}

where $a$, $b$, and $c$ are coefficients that can either be calculated using the perturbation theory or obtained from a quadratic fit of the measured mass excesses of the multiplet members.  The IMME can thus be used to predict the energies of unobserved multiplet states, and measurements of these states can be compared with the quadratic form of the IMME in order to test its validity.  A breakdown of the IMME could indicate a failure of the perturbation theory and a need for higher-order terms, the presence of many-body charge-dependent forces \cite{JaneckeNPA1969}, isospin mixing of the IAS with other nearby states of different isospin \cite{SignoracciPRC2011}, or inaccurate measurements.

Historically, the IMME has been very successful at describing experimental values, requiring very few deviations from the quadratic form.  As discussed in Refs. \cite{BritzADNDT1998} and \cite{LamADNDT2013}, in situations where the fit of the quadratic form is very poor, a cubic or quartic form with extra terms $dT_z^3$ or $eT_z^4$ may be required.  Typically, these terms have been determined empirically to be either very small, $\lesssim$1 keV, or consistent with zero.  A number of situations where a $d$ term has been required are noted in Ref. \cite{MacCormickNPA2014}, including the $A = 8$ and $A = 32$ $T = 2$ quintets and the $A = 9$ and $A = 35$ $T = 3/2$ quartets; these and other cases are discussed here.

In the $A = 8, T = 2$ quintet mentioned above, multiple studies \cite{BritzADNDT1998,CharityPRC2011} have noted the need for a significant cubic term in the IMME, $d = 7.4(14)$ and $11.1(23)$ keV for Refs. \cite{BritzADNDT1998} and \cite{CharityPRC2011}, respectively.  The recent evaluation of Ref. \cite{MacCormickNPA2014} confirmed that a quartic function was most likely an even better fit to the data and suggested the need for both theoretical and experimental studies of the multiplet members to address the IMME breakdown.  A Penning trap measurement of multiple isotopes including \A{21}Mg \cite{GallantPRL2014} reported breakdowns of the IMME for the $J^\pi = 5/2^+$ and $J^\pi = 1/2^+$ $A = 21, T = 3/2$ quartets, requiring cubic terms of $d = 6.7(13)$ keV and $-4.4(14)$ keV, respectively.  The $A = 32, T = 2$ quintet is currently the most precisely measured quintent. Here, a precise Penning trap measurement of \A{32}Si \cite{KwiatkowskiPRC2009} led to an observed breakdown of the IMME, requiring a small, but very significant, cubic term $d = 1.00(9)$ keV, which was supported by a measurement of the \A{32}Cl mass using the \A{32}S(\A{3}He,$t$)\A{32}Cl reaction \cite{WredePRC2010}.  A later precision measurement of the \A{31}S mass \cite{KankainenPRC2010} led to an even greater precision on the \A{32}Cl mass excess and found that no combination of various literature values could produce a fit that validated the IMME for the $A = 32, T = 2$ quintet.  A recent review of mass measurements \cite{KankainenEPJA2012} suggested that new mass measurements of other multiplet members might revalidate the IMME in this quintet, and a theoretical study of the quintet \cite{SignoracciPRC2011} demonstrated that the IMME deviation, as well as the observed isospin-forbidden proton decay from the $T = 2$ \A{32}Cl IAS and a correction to the $0^+ \rightarrow 0^+$ superallowed decay from \A{32}Ar, could be traced to isospin mixing of the $T = 2$ states with $T = 1$ states.  In the $A = 35, T = 3/2$ case, a relatively large $d$ coefficient of $-3.39(41)$ keV was clearly required to fit the data.  Although no definite solution has been found yet, both inaccurate experimental data and isospin mixing have been suggested as potential causes for the breakdown \cite{YazidjianPRC2007}.

In several instances of IMME breakdown, additional study has revalidated the quadratic IMME.  Reference \cite{GallantPRL2014} reported, in addition to the findings on \A{21}Mg, a new \A{20}Mg mass and a resulting IMME breakdown requiring a cubic term of 2.8(11) keV for the $A = 20, T = 2$ quintent.  In this case, a recent experimental measurement of \A{20}Mg $\beta$ decay \cite{GlassmanPRC2015} used the superallowed $0^+ \rightarrow 0^+$ transition to the $T = 2$ \A{20}Na IAS and the state's subsequent $\gamma$ decay to deduce an excitation energy for the IAS. This result was 28 times more precise than the previous measurement and, together with the ground state mass excess of \A{20}Na, was shown to revalidate the IMME for the $A = 20, T = 2$ quintet.  A storage ring mass measurement of a number of $fp$ shell nuclei, including \A{53}Ni \cite{ZhangPRL2012}, found that an IMME fit of the lowest $A = 53, T = 3/2$ quartet required an enormous cubic term of $d = 39(11)$ keV, a $3.5\sigma$ deviation from the quadratic IMME.  In this latter case, as in the $A = 20$ case, the IMME was revalidated after a measurement of \A{53}Ni $\beta$-delayed $\gamma$ decay \cite{SuPLB2016} which produced a more precise \A{53}Co IAS excitation energy and a cubic IMME fit with a $d$ coefficient compatible with zero.  In the $A = 9, T = 3/2$ case mentioned above, a relatively large $d$ coefficient of $6.33(164)$ keV was required to fit the IMME.  This anomaly was found through high-precision mass measurements of \A{9}Li and \A{9}Be to be the result of isospin mixing in \A{9}B and \A{9}Be \cite{BrodeurPRL2012}.

In the lowest $A = 31$, $T = 3/2$ quartet, it has until recently been difficult to test the IMME because the experimental mass excess value of \A{31}Cl has been relatively imprecise.  A 1977 experimental measurement of the \A{36}Ar(\A{3}He,\A{8}Li)\A{31}Cl $Q$ value resulted in a mass excess value of $\Delta = -7070 \pm 50$ keV \cite{benensonPRC1977}.  Subsequent evalutions of the IMME \cite{BenensonRMPhys1979,BritzADNDT1998,LamADNDT2013,MacCormickNPA2014} have included adjusted central values of this mass excess, but the uncertainty has remained.  In contrast to high-precision mass excess and excitation energy values of the relevant states in \A{31}S \cite{MoalemPRC1973,KankainenPRC2010,KankainenEPJA2006} and \A{31}P \cite{RedshawPRL2008,NDS_A31_2013} (based on mass measurements of those nuclei and experimental measurements of their excitation energies)  and in \A{31}Si (based on mass measurements of \A{29}Si \cite{RainvilleNature2005} and neutron-capture reaction measurements linking the isotopes from \A{29}Si to \A{31}Si \cite{IslamPRC1990,RamanPRC1992,RottgerIMIT1997}), the 50-keV uncertainty in the \A{31}Cl mass excess has hindered attempts to test the IMME stringently in the lowest quartet.  A recent Penning trap mass measurement of \A{31}Cl finally obtained a value for the ground state mass excess 15 times more precise than previous estimates \cite{KankainenPRC2016}, leading to an IMME breakdown in the lowest quartet; the IMME fit required an unusually large cubic term, with $d = -3.5(11)$ keV. 

Similar to the lowest quartet, uncertainties associated with both the energy of the first excited state in \A{31}Cl and the identity of the second $T = 3/2$ state in \A{31}S have precluded a quality test of the IMME in the second quartet.  In fact, a tentative measurement of the first \A{31}Cl excited state via \A{31}Ar $\beta$ decay \cite{AxelssonNPA1998} was the only evidence for the observation of that state \cite{WredePRC2009} until a recent Coulomb-breakup experiment was performed to confirm the existence of the state \cite{LangerPRC2014}.  The excitation energy was found in Ref. \cite{LangerPRC2014} to be $E_x = 782(32)$ keV, leaving the identity of the second $T = 3/2$ state in \A{31}S as the primary ambiguity in the quartet.  Various sources have reported excitation energies for the \A{31}S state ranging from a definite $T = 3/2$ assignment for a state at $E_x = 7006(25)$ keV using the \A{29}Si(\A{3}He,$n$)\A{31}S reaction \cite{DavidsonNPA1975} with somewhat low precision to a relatively precise, but tentative, assignment for a state at $E_x = 7036(2)$ keV \cite{Wrede2009PRCnumber2}, with alternative candidates at 6975(3) \cite{BrownPRC14,Wrede2009PRCnumber2} and 7053(2) keV \cite{Wrede2009PRCnumber2}.  

Although this $J^\pi = 1/2^+$ \A{31}S state is expected to be nearly 1 MeV above the proton threshold, the proton emission is isospin forbidden and, therefore, it should have a substantial $\gamma$-decay branch unlike the other low-spin levels in the region \cite{GlassmanPRC2015}.  Precise observation of a high energy $\gamma$-ray transition from a low-spin state in this region would be a signature of the second $T = 3/2$ state, allowing for a precise determination of its energy.  The shell model predicts that the state decays predominantly to the ground state, and shell model calculations using the universal $sd$-shell version ``B'' (USDB) \cite{RichterPRC2008} and the recently-developed version ``E'' (USDE) \cite{BennettPRL2015} models predict a \A{31}S state 745(50) keV above the \A{31}S IAS energy of $E_x = 6279$ keV.  In the shell model, this state has a \A{31}Cl $\beta$ feeding of 0.03(2)\% and a ground-state $\gamma$-decay branch of $\Gamma_\gamma^{g.s.} / \Gamma_\gamma = 0.95(4)$.  

The present paper reports the results from a \A{31}Cl $\beta$-decay study and presents potential solutions to the problem of IMME breakdown for the lowest $A = 31, T = 3/2$ quartet based on the observation of isospin mixing in \A{31}S.  In addition, a precision measurement of the second $T = 3/2$ \A{31}S state is reported, allowing for the most stringent test of the IMME to date for the second $A = 31, T = 3/2$ quartet.

\section{Experiment}

The present experiment is one in a series of recent $\beta$-delayed $\gamma$-decay experiments to investigate the $sd$ shell using fast neutron-deficient beams at the National Superconducting Cyclotron Laboratory \cite{BennettPRL2013,SchwartzPRC2015,perezloureiroPRC2016,BennettPRL2015,GlassmanPRC2015}.  In particular, the $\beta$-delayed $\gamma$ decay of \A{31}Cl was measured using an experimental procedure that was already described in \cite{BennettPRL2015}.  Briefly, a fast beam of up to 9000 \A{31}Cl ions per second was implanted into a plastic scintillator, which acted as a $\beta$-decay trigger.  The $\beta$-delayed $\gamma$ rays were detected using the Clovershare array: nine ``clover'' detectors of four Ge crystals each, surrounding the plastic scintillator.  Data from the crystals were gain-matched and calibrated to produce ${\beta}{\gamma}$ and ${\beta}{\gamma}{\gamma}$ spectra, the latter of which were gated on a variety of deexciting $\gamma$ rays.  From these data, a decay scheme was constructed including the observed \A{31}S levels and their excitation energies and $\beta$ feedings.  Absolute ${\beta}{\gamma}$ intensities for the observed $\gamma$ rays were also determined.  In the present work we focus on the $T = 3/2$ states.

\section{Results and Discussion}

\subsection{First $A = 31, T = 3/2$ Quartet}

\subsubsection{Experimental Results}

We observed the $\gamma$-ray deexcitation of a \A{31}S state at $E_x = 6390.2(7)$ keV, as previously reported in Ref. \cite{BennettPRL2015}.  Neither the $\beta$ feeding nor the $\gamma$-decay branching of the state match our USDB predictions \cite{RichterPRC2008} without isospin mixing, and the state's $\beta$ feeding was abnormally high for a state at such a high energy.  By computing the Fermi strengths $B$ for both the $T = 3/2$ \A{31}S IAS at 6279.0(6) keV and this state, it was discovered that the two states were mixing isospin strongly.  The mixing of the state at 6390 keV allowed for an unambiguous spin and parity identification of $J^\pi = 3/2^+$.  The positive identification of the state has implications for the \A{30}P\pg\A{31}S reaction rate in the astrophysical environment of a classical nova outburst; these findings are discussed in Ref. \cite{BennettPRL2015}. Excitation energies, $\gamma$-decay energies, and $\beta$ feedings of the two states are summarized in Table \ref{energies}.  

\subsubsection{IMME}

Here we explore the impact of isospin mixing on the IMME for the lowest $A = 31, T = 3/2$ quartet.  In order to use the IMME fit, we needed values for both the ground-state mass excesses of the multiplet members and the \A{31}S and \A{31}P IAS excitation energies.  For the lowest quartet, we used the values in Ref. \cite{KankainenPRC2016} for \A{31}Cl, \A{31}P, and \A{31}Si.  For \A{31}S, we used the value for the \A{31}S IAS excitation energy obtained from the present work \cite{BennettPRL2015} rather than the value of $E_x = 6280.60(16)$ keV used in Ref. \cite{KankainenPRC2016}, which is from the $A = 31$ Nuclear Data Sheets (NDS) \cite{NDS_A31_2013}.  The value in Ref. \cite{NDS_A31_2013} is based on a fit of gamma-ray energies from a measurement of \A{31}Cl beta decay \cite{SaastamoinenThesis}.  However, since the NDS value does not factor in the 1.5-keV systematic uncertainty reported in Ref. \cite{SaastamoinenThesis}, we considered it to be less precise than the value obtained in the present work \cite{BennettPRL2015}, which includes both statistical and systematic uncertainty.  Nevertheless, it is worthwhile to note that the excitation energy value from the present work is consistent with the value from Refs. \cite{NDS_A31_2013,SaastamoinenThesis} when systematic uncertainties are included.

As discussed in Ref. \cite{KankainenPRC2016}, with the new high-precision \A{31}Cl mass excess, a fit of the quadratic IMME fails, requiring a large coefficient for the cubic term, $d = -3.5(11)$ keV.  Using our value of 6279.0(6) keV for the observed IAS excitation energy, the quadratic fit also fails, requiring a coefficient for the cubic term of $d = -4.3(11)$ keV.  This failure of the quadratic fit is independent of whether we use our value for the \A{31}S excitation energy or the value from Ref. \cite{SaastamoinenThesis}.  The inputs and outputs of this fit are reported in Tables \ref{badfit1} and \ref{badfit2}, respectively.

The authors of Ref. \cite{KankainenPRC2016} hypothesize that isospin mixing could help explain the observed IMME breakdown.  In the case where two states mix, the following equations may be used to calculate the empirical mixing matrix element and unperturbed level spacing:

\begin{equation}
E =  \sqrt{D^2 + 4 V^2}
\label{1}
\end{equation}
\begin{equation}
E = D + 2\delta
\label{3}
\end{equation}
\begin{equation}
R = \sqrt{\frac{B_2}{B_1}}
\label{4}
\end{equation}
\begin{equation}
D = E~\dfrac{1 - R^2}{1 + R^2}
\label{2}
\end{equation}
\begin{equation}
V = E~\dfrac{R}{1 + R^2}
\label{3}
\end{equation}

where $E$ is the observed spacing, $D$ is the unperturbed level spacing, $V$ is the mixing matrix element, $\delta$ is the perturbation, and $R$ may be calculated from the Fermi strengths $B$ of $\beta$-decay transitions to the two mixed states (Eq. \ref{4}).  With these equations and the Fermi strengths calculated in Ref. \cite{BennettPRL2015} [$B_1 = 2.4(1)$ and $B_2 = 0.48(3)$], we derive an empirical mixing matrix element and unperturbed level spacing of 41(1) and 74(2) keV, respectively.  Using this unperturbed level spacing and the observed energies of the two states, we calculate the unperturbed energy of the IAS to be 6297.6(13) keV. 

\begin{figure}
\includegraphics[width=\columnwidth]{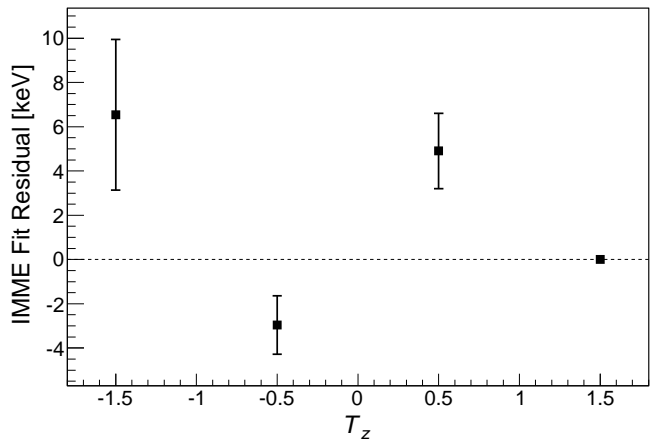} %%,natwidth=1748,natheight=5171   ,height=0.8\textheight
\caption{Residuals for the quadratic IMME fit of the lowest $A = 31, T = 3/2$ quartet (Tables \ref{inputTable1} and \ref{outputTable1}) after accounting for the observed isospin mixing in \A{31}S.}
\label{firstquartet}
\end{figure}

In order to test the hypothesis that isospin mixing is affecting the quadratic IMME fit in the lowest $A = 31, T = 3/2$ quartet, we tried fitting the IMME including this unperturbed \A{31}S IAS energy along with the new \A{31}Cl mass measured in Ref. \cite{KankainenPRC2016} and the other, precisely known values for \A{31}Si and \A{31}P \cite{AME_2012}.  However, this does little to fix the breakdown problem: the reduced chi-squared value of the quadratic fit actually increases from $\chi^2 / \nu = 11.6$, in Ref. \cite{KankainenPRC2016}, and 16.0, using the new observed IAS $E_x$, to 17.0. For the cubic fit, the coefficient $d$ also becomes larger in magnitude, changing from $-3.5(11)$ keV or $-4.3(11)$ keV to $+5.0(12)$ keV.  Input energies for the unperturbed-energy fits from the present work are listed in Table \ref{inputTable1}, and fit output parameters for the quadratic and cubic fit are listed in Table \ref{outputTable1}.  Residuals for the quadratic fit are shown in Fig. \ref{firstquartet}.

\begin{table}
\caption{\label{energies} Experimentally determined excitation energies $E_x$, $\gamma$-decay energies $E_\gamma$, and \A{31}Cl $\beta$-decay feedings $I_{\beta^+}$ for the first two $T = 3/2$ \A{31}S states, as well as the 6390-keV state that mixes with the first $T = 3/2$ state.}

\begin{ruledtabular}
\begin{tabular}{c c c c c}
 $ T $ 	&	$J^\pi$	& $E_x$ (keV)		& $E_\gamma$ (keV)		& $I_{\beta^+}$	(\%)							\\
 \hline                                                                             \\
 $3/2$	&	$3/2^+$	& $6279.0(6)$		& $6278.4(6)$	 				& $18.7(9)$					\\
 $1/2$	&	$3/2^+$	& $6390.2(7)$   & $6389.5(7)$  				& $3.38(16)$       \\
 $3/2$	&	$1/2^+$	& $7050.0(8)$   & $7049.2(8)$   			& $0.047(5)$      \\

\end{tabular}
\end{ruledtabular}
\end{table}

\begin{table}
\caption{\label{badfit1} Ground-state mass excess $\Delta$ and excitation energy $E_x$ values used as input for the IMME fits of the lowest $A = 31, T = 3/2$ quartet. Except for the observed excitation energy of the \A{31}S IAS, which is from Ref. \cite{BennettPRL2015}, all values are the same as in Ref. \cite{KankainenPRC2016}.}

\begin{ruledtabular}
\begin{tabular}{c c c c}
 $  $ 	Nucleus		& $ T_z $										& $\Delta$ (keV)								&	$E_x$ (keV) \\
 \hline          \\                                                             
 \A{31}Cl         & $-3/2$										& $-7034.7(34)$									& $0$							\\
 \A{31}S					& $-1/2$                    & $-19042.52(23)$               & $6279.0(6)$				\\
 \A{31}P   				& $+1/2$                    & $-24440.5411(7)$              & $6380.8(17)$		\\
 \A{31}Si         & $+3/2$                    & $-22949.04(4)$                & $0$							\\

\end{tabular}
\end{ruledtabular}
\end{table}

\begin{table}
\caption{\label{badfit2} Output coefficients for the quadratic and cubic IMME fits for the lowest $A = 31, T = 3/2$ quartet using input data from Table \ref{badfit1}.  All coefficient values are in units of keV.  The cubic fit did not contain any degrees of freedom, so the $\chi^2 / \nu$ value is undefined and hence ommitted. }

\begin{ruledtabular}
\begin{tabular}{c c c}
 $  $ 						& Quadratic									& Cubic								\\
 \hline                                                               \\
 $a$				        & $-15466.3(9)$						& $-15464.1(10)$					\\
 $b$				 				& $-5302.4(10) $         	& $-5295.2(20)$       \\
 $c$				   			& $209.2(9)$              & $209.9(10)$      \\
 $d$        				& $    $                  & $-4.3(11)$        \\
$\chi^2 / \nu$		&	$ 16.0 / 1$							&	$    $											\\

\end{tabular}
\end{ruledtabular}
\end{table}

\begin{table}
\caption{\label{inputTable1} Ground-state mass excess $\Delta$ and excitation energy $E_x$ values used as input for the IMME fits of the lowest $A = 31, T = 3/2$ quartet. Except for the unperturbed excitation energy of the \A{31}S IAS, which is from the present work \cite{BennettPRL2015}, all values are the same as in \cite{KankainenPRC2016}.}

\begin{ruledtabular}
\begin{tabular}{c c c c}
 $  $ 	Nucleus		& $ T_z $										& $\Delta$ (keV)								&	$E_x$ (keV) \\
 \hline          \\                                                             
 \A{31}Cl         & $-3/2$										& $-7034.7(34)$									& $0$							\\
 \A{31}S					& $-1/2$                    & $-19042.52(23)$               & $6297.6(13)$				\\
 \A{31}P   				& $+1/2$                    & $-24440.5411(7)$              & $6380.8(17)$		\\
 \A{31}Si         & $+3/2$                    & $-22949.04(4)$                & $0$							\\

\end{tabular}
\end{ruledtabular}
\end{table}

\begin{table}
\caption{\label{outputTable1} Output coefficients for the quadratic and cubic IMME fits for the lowest $A = 31, T = 3/2$ quartet using input data from Table \ref{inputTable1}.  All coefficient values are in units of keV.}

\begin{ruledtabular}
\begin{tabular}{c c c}
 $  $ 						& Quadratic									& Cubic								\\
 \hline                                                               \\
 $a$				        & $-15453.0(12)$								& $-15454.6(12)$					\\
 $b$				 				& $-5307.0(10) $         		  & $-5316.1(24)$       \\
 $c$				   			& $206.4(10)$              & $205.2(10)$      \\
 $d$        				& $    $                    & $5.0(12)$        \\
$\chi^2 / \nu$		&	$ 17.0 / 1$							&	$    $											\\

\end{tabular}
\end{ruledtabular}
\end{table}

Given the observed isospin mixing in \A{31}S, it is likely that there is similar mixing present in the mirror nucleus \A{31}P, which has not yet been directly observed.  Using the unperturbed energy of the \A{31}S IAS from Ref. \cite{BennettPRL2015} and the results of Ref. \cite{KankainenPRC2016}, the IMME can be used to predict an ``unperturbed" energy for the lowest $T = 3/2$ state in \A{31}P of 6390.8(24) keV, only $10$ keV higher than the current energy value of the \A{31}P IAS, $E_x = 6380.8(17)$ keV.  If this state, like the \A{31}S IAS, mixes isospin with a nearby higher-energy $T = 1/2$ state, its unperturbed energy could be high enough to revalidate the IMME for the quartet after accounting for the isospin mixing.

\begin{figure}
\includegraphics[width=\columnwidth]{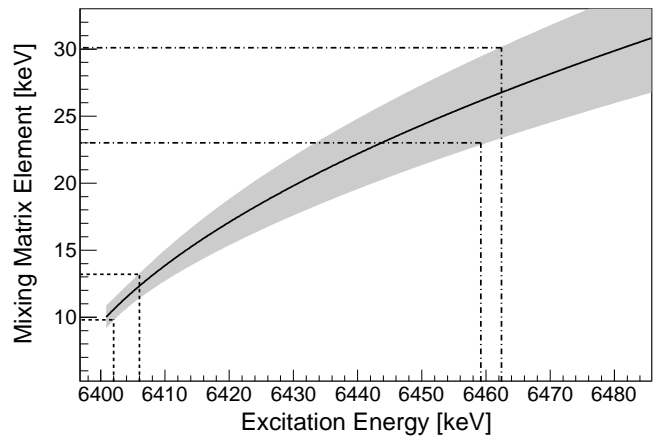} %%,natwidth=1748,natheight=5171   ,height=0.8\textheight
\caption{Isospin mixing matrix element, including $1\sigma$ confidence band, of a hypothetical state engaged in isospin mixing with the \A{31}P IAS at 6381 keV as a function of the observed excitation energy of the second state.  The band is derived under the assumption that the IMME provides a good fit of the data after accounting for isospin mixing.  The dotted (left) and dot-dashed (right) lines show the 1$\sigma$ bounds obtained using this prediction when the USD mixing matrix element and 6461-keV state energy, respectively, are used as inputs.}
\label{predict31P}
\end{figure}

At first glance, however, it appears that no such state is known to exist experimentally.  No nearby higher-energy states listed in the 2013 $A = 31$ Nuclear Data Sheets \cite{NDS_A31_2013} have the same spin and parity ($J^\pi = 3/2^+$) as the \A{31}P IAS.  It is possible, however, to derive combinations of excitation energy and mixing matrix element for such a state that would revalidate the quadratic IMME.  Using Eqs. \ref{1} and \ref{2} along with the observed and predicted energies of the lowest $T = 3/2$ \A{31}P state and solving for $V$, the result is a curve for $E_x \geq 6401$ keV  (Fig. \ref{predict31P}).  The lowest energy solution at 6401 keV corresponds to two degenerate unperturbed states at $E_x \approx 6391$ keV, perturbed by $\pm$ 10 keV by mixing.  

As a naive empirical prediction, the assumption that the unperturbed energy spacing is identical in this case to the \A{31}S case ($74(2)$ keV) yields a second state at $E_x = 6464.8(35)$ keV, with an associated mixing matrix element of 27.2(35) keV.  Coincidentally, this predicted energy is near a known \A{31}P state at 6460.8(16) keV, listed as $J^\pi = 5/2^+$ in Ref. \cite{NDS_A31_2013}.  Although some sources \cite{EndtNPA1990} have committed to a definite spin and parity assignment for this state, multiple experimental studies \cite{WolffNPA1970,AlJadirJPGNP1980, McCullochNPA1984,VernottePRC1990,VernotteNPA1999}, while potentially favoring the $5/2^+$ assignment, have not excluded a $3/2^+$ assignment.  Further, as noted in Ref. \cite{NDS_A31_2013}, another study \cite{KaschlNPA1969} has even labeled the state as $J^\pi = 1/2^+$, further complicating the matter of its spin and parity.  If the state did in fact have $J^\pi = 3/2^+$, it could mix with the IAS at 6381 keV.

\subsubsection{Shell Model}

In order to facilitate the search for the hypothetical state mixing with the \A{31}P IAS, we have used both USDE and USDB to predict energy levels and mixing matrix elements for the \A{31}P IAS and its nearby states.  As in the \A{31}S case, both USDE and USDB predict a triplet of $J^\pi = 3/2^+$ states involved in mixing.  The results of both the USDE and USDB calculation are reported in Table \ref{USDCalcs}.  The mixing matrix element values obtained are substantially smaller than those for \A{31}S \cite{BennettPRL2015}.  While there are experimental candidates for the lower state in the triplet at 6233 keV and 6158 keV \cite{NDS_A31_2013}, again no higher state in the vicinity is immediately apparent.  The closest candidate is the state previously mentioned at 6461 keV, but as it requires a relatively high mixing matrix element (27(3) keV) compared to theory (Fig. \ref{predict31P}), it should be regarded as a tentative solution at best.  Consequently, experimental searches for additional levels are needed to test the likelihood that the 6461-keV state is the $3/2^+$ state mixing with the \A{31}P IAS and to find other potential states which could fulfill that role.  The shell-model matrix elements for the mixed states (Table \ref{USDCalcs}) and the functional form in Fig. \ref{predict31P} may be used to predict the energy region in which the mixed state is likely to exist: the theoretical upper and lower bounds are $\approx 6402$ and $\approx$ 6406 keV, respectively.  Searches for the hypothetical mixing state should thus focus on the region slightly above $E_x = 6400$ keV.

\begin{table}
\caption{\label{USDCalcs} Calculated excitation energies $E_x$ and mixing matrix elements $V$ of the triplet of isospin-mixed states including the lowest $T = 3/2$ state in \A{31}P for both USDB and USDE interactions.  The matrix elements listed are between the listed $T = 1/2$ state and the $T = 3/2$ state.  All values are in units of keV.}

\begin{ruledtabular}
\begin{tabular}{c c c c c c}
 $  $ 						& $J^\pi$	& USDB $E_x$			&	USDB $V$					& USDE $E_x$				&		USDE $V$			\\
 \hline                                      																         \\
 $E_1 (T = 1/2)$	& $3/2^+$	& $6258$					&	$8.3$							& $6118$						&		$4.2$					\\
 $E_2 (T = 3/2)$	& $3/2^+$	&$6364$  				& $ $      				& $6236$ 						&		$ $      \\
 $E_3 (T = 1/2)$	& $3/2^+$	&$6579$					&	$10.9$						& $6383$  					&		$12.7$   \\
 
\end{tabular}
\end{ruledtabular}
\end{table}

\begin{figure}
\includegraphics[width=\columnwidth]{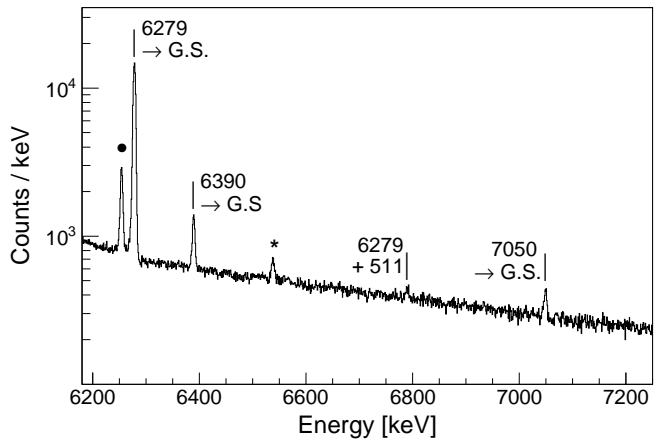} %%,natwidth=1748,natheight=5171   ,height=0.8\textheight
\caption{High-energy portion of the \A{31}Cl $\beta$-coincident $\gamma$ spectrum obtained in Ref. \cite{BennettPRL2015}.  The transitions from the IAS at 6279 keV, the state at 6390 keV, and the $J^\pi = 1/2^+$ \A{31}S state at 7050 keV to the ground state are all labeled with a vertical line.  The peak marked with an asterisk at 6539 keV is the first escape peak corresponding to the 7050-keV transition.  The peak at 6791 keV is the sum peak between the strong 6280-keV photopeak \cite{BennettPRL2015} and the 511 keV annihilation photopeak.  The peak at 6255 keV is a photopeak corresponding to a transition from a $T = 1/2$ \A{31}S state to the ground state.}
\label{spec7050}
\end{figure}

\begin{figure}
\includegraphics[width=\columnwidth]{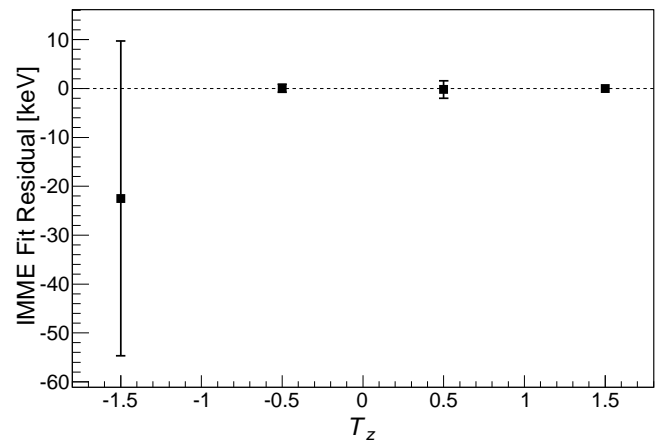} %%,natwidth=1748,natheight=5171   ,height=0.8\textheight
\caption{Residuals for the quadratic IMME fit of the second-lowest $A = 31, T = 3/2$ quartet (Tables \ref{inputTable2} and \ref{outputTable2}).}
\label{residuals2}
\end{figure}

\begin{table}
\caption{\label{inputTable2} Ground-state mass excess $\Delta$ and excitation energy $E_x$ values \cite{NDS_A31_2013} used as input for the IMME fits of the second-lowest $A = 31, T = 3/2$ quartet.}

\begin{ruledtabular}
\begin{tabular}{c c c c}
 $  $ 	Nucleus		& $ T_z $										& $\Delta$ (keV)								&	$E_x$ (keV) \\
 \hline          \\                                                             
 \A{31}Cl        & $-3/2$		& $-7034.7(34)$			& $782(32)$							\\
 \A{31}S	  & $-1/2$                    & $-19042.52(23)$               & $7050.0(8)$				\\
 \A{31}P   	  & $+1/2$                    & $-24440.5411(7)$              & $7141.1(18)$		\\
 \A{31}Si         & $+3/2$                    & $-22949.04(4)$                & $752.23(3)$							\\

\end{tabular}
\end{ruledtabular}
\end{table}

\begin{table}
\caption{\label{outputTable2} Output coefficients for the quadratic and cubic IMME fits for the second-lowest $A = 31, T = 3/2$ quartet using input data from Table \ref{inputTable2}.  All coefficient values are in units of keV.}

\begin{ruledtabular}
\begin{tabular}{c c c}
 $  $ 			& Quadratic			& Cubic								\\
 \hline                                                               \\
 $a$			& $-14697.3(14)$		& $-14698.6(22)$					\\
 $b$			& $-5307.2(19) $    & $-5306.0(25)$       \\
 $c$			& $205.0(18)$       & $211(8)$      \\
 $d$      & $    $            & $-4(5)$        \\
$\chi^2 / \nu$	& $ 0.51 / 1$		&	$    $											\\

\end{tabular}
\end{ruledtabular}
\end{table}

\subsection{Second $A = 31, T = 3/2$ Quartet}

\subsubsection{Experimental Results}

An isolated $\gamma$-ray peak corresponding to a laboratory energy $E_\gamma = 7049.2(8)$ was observed in our $\beta\gamma$ spectrum (Fig. \ref{spec7050}), 770 keV above the IAS, as predicted for the second $T = 3/2$ \A{31}S state by our shell model calculations.  It did not appear in any of our $\beta\gamma\gamma$ coincidence spectra, so the simplest interpretation is that this transition is from a \A{31}S level at $E_x = 7050.0(8)$ keV undergoing a transition to the ground state.  The $\beta$ feeding of 0.047(5)\% for this state is consistent with shell model predictions, and no other $\gamma$-ray transitions de-exciting this state were observed, implying that it decays predominantly to the ground state.  The agreement of the state's excitation energy and $\beta$ feeding with the shell-model prediction, its singular $\gamma$ branch to the ground state, and a small observed $\beta$-$p$ branch \cite{WredePRC2009} all provide evidence that it is indeed the second $T = 3/2$ state, with $J^\pi = 1/2^+$, in \A{31}S.  

\subsubsection{IMME} 

A quadratic fit of the IMME using the observed 7050-keV state energy and the energies of the other three quartet members results in a good fit with $\chi^2 / \nu = 0.51 / 1$ and a $p$ value of 0.48.  This is further confirmation that the \A{31}S state at 7050 keV is the \A{31}S member of the second $T = 3/2, A = 31$ quartet.  Input mass excesses and excitation energies are reported in Table \ref{inputTable2}.  Output parameters for both the quadratic and cubic fits are reported in Table \ref{outputTable2}.  Residuals for the quadratic fit of all four states are shown in Fig. \ref{residuals2}.

\begin{table}
\caption{\label{USDCalcs2} Calculated excitation energies $E_x$ and mixing matrix elements $V$ of the triplets of states involved in mixing with the second-lowest $T = 3/2$ states in both \A{31}S and \A{31}P, for both USDB and USDE interactions.  The matrix elements listed are between the listed $T = 1/2$ state and the $T = 3/2$ state.  All values are in units of keV.}

\begin{ruledtabular}
\begin{tabular}{c c c c c c}
	
 \A{31}S 					& $J^\pi$		& USDB $E_x$			&	USDB $V$					& USDE $E_x$				&		USDE $V$			\\
 \hline                                      																         \\
 $E_1 (T = 1/2)$	& $1/2^+$		&	$7234$					&	$7.8$							& $6421$						&		$6.8$					\\
 $E_2 (T = 3/2)$	& $1/2^+$		& $7271$  				& $ $      				& $6944$ 						&		$ $      \\
 $E_3 (T = 1/2)$	& $1/2^+$		& $7814$					&	$22$							& $7117$  					&		$9.4$   \\
 \hline
 \hline
 \\
	\A{31}P 				& $J^\pi$		& USDB $E_x$			&	USDB $V$					& USDE $E_x$				&		USDE $V$			\\
 \hline                                                  																         \\
 $E_1 (T = 1/2)$	& $1/2^+$		& $7251$					&	$6.5$						& $6417$					&		$3.1$					\\
 $E_2 (T = 3/2)$	& $1/2^+$		& $7310$  				& $ $      			& $6982$ 					&		$ $      \\
 $E_3 (T = 1/2)$	& $1/2^+$		& $7861$					&	$6.1$						& $7127$  					&		$7.2$   \\

\end{tabular}
\end{ruledtabular}
\end{table}

Although the IMME fit is very good with the measured mass excesses and excitation energies for the second quartet members, it is possible that a small amount of undetected isospin mixing occurs, similar to the mixing in the lowest quartet.  While potential candidate $T = 1/2$ states exist for mixing in each of these nuclei, no experimental evidence was observed to positively identify such a state in \A{31}S.  For example, no $\gamma$-ray transitions were observed for any states between $E_x = 6400$ keV and $E_x = 7050$ keV in either the ${\beta}{\gamma}$ or the ${\beta}{\gamma}{\gamma}$ coincidence spectra.

\subsubsection{Shell Model}

To estimate the amount of mixing that might occur, we have used both the USDE and USDB calculations to predict energy levels and mixing matrix elements for both \A{31}S and \A{31}P (Table \ref{USDCalcs2}).  Both models produce small mixing matrix elements, consistent with the implication from our IMME fit that the mixing is small, with the lack of observation of other $\gamma$-ray branches in the energy region, and with the small ratio of proton emission to $\gamma$ decay of the 7050-keV state.

\subsubsection{Prediction of \A{31}Cl First Excited State Energy}

Using our high-precision measurement of the excitation energy of the second $T = 3/2$ state in \A{31}S, it is possible to predict the energy of the first excited \A{31}Cl state with a higher precision.  Using the \A{31}S, \A{31}P, and \A{31}Si input mass excess values and excitation energies in Table \ref{inputTable1} to produce the IMME curve, and accounting for the uncertainty introduced by the possibility of isospin mixing via the $d$ coefficient in the cubic fit, the resulting IMME mass excess is $\Delta = -6276(10)$ keV.  When combined with the known \A{31}Cl ground state mass excess from Ref. \cite{KankainenPRC2016} and its uncertainty, the predicted excitation energy of the state is $E_x = 759(11)$ keV, consistent with the measured value of $E_x = 782(32)$ keV  \cite{LangerPRC2014}.  It is also possible to calculate the energy of the state using the \A{30}S $+ p$ resonance energy based on the $\beta$$p$ measurement, $E_r = 461(15)$ \cite{WredePRC2009, AxelssonNPA1998} and the recent value of the proton separation energy, $S_p = 265(4)$ \cite{KankainenPRC2016}: The result is $E_x = 726(16)$, which is consistent with our prediction within 1.8 combined standard deviations and with the value from Ref. \cite{LangerPRC2014} within 1.6 combined standard deviations.  Given the slight tension between the value based on the ${\beta}{p}$ measurement \cite{AxelssonNPA1998} and the other two values, a new measurement of \A{31}Ar $\beta$ decay \cite{KoldstePRC2013,KoldstePRC2014} with high sensitivity to low-energy protons would be an interesting study.

\section{Conclusions}

The problem of IMME breakdown in the lowest $A = 31, T = 3/2$ quartet may be a result of isospin mixing in one or more of the multiplet members.  We have measured the excitation energy of the \A{31}S IAS and found that it is isospin mixed with a nearby $T = 1/2$ state.  However, we have shown that the isospin mixing in \A{31}S alone cannot explain the IMME breakdown observed in Ref. \cite{KankainenPRC2016}, but that incorporating similar mixing in \A{31}P could account for the breakdown.  Better experimental data are needed to search for the hypothesized $J^\pi = 3/2^+, T = 1/2$ \A{31}P state that may be mixing with the IAS.  The state at 6461 keV is the best existing candidate for this state, but is not consistent with shell model predictions and should be investigated further.  It is not certain that every \A{31}P state in the energy region has been observed.  Future experiments could focus on gamma spectroscopy of this region in \A{31}P to complement the numerous charged particle reaction measurements carried out to-date.  The question of isospin mixing in \A{31}P likely holds the key to explaining the IMME breakdown in the lowest $A = 31, T = 3/2$ quartet; uncovering the structure in the important energy region slightly above $E_x = 6400$ keV could lead to a deeper understanding of the perturbative effects of isospin mixing on this widely-used theoretical model.

In addition, the clear identification of the second $ T = 3/2$ \A{31}S state at 7050 keV in the present work has completed the second $A = 31, T = 3/2$ quartet.  Three of the four quartet member states now have mass-excess uncertainties $<2$ keV and their masses are well-described by the quadratic form of the IMME. Further studies of this multiplet could focus on reducing the uncertainty of the energy of the first excited \A{31}Cl state.  A new \A{31}Ar $\beta$ decay study could provide an independent measurement of the $\beta$-delayed proton energy, and in-beam $\gamma-$ray spectroscopy, while challenging due to the very small expected $\gamma$-decay branching ratio of the state, might present a novel approach to measuring the excitation energy of this state.

The researchers gratefully acknowledge the dedicated effort of the NSCL operations staff to ensure the delivery of multiple very pure beams. This work was supported by the U.S. National Science Foundation under Grants No. PHY-1102511, PHY-1404442, PHY-1419765, and PHY-1431052, the US Department of Energy, National Nuclear Security Administration under Grant No. DE-NA0000979, and the Natural Sciences and Engineering Research Council of Canada.  We also gratefully acknowledge use of the Yale Clovershare array.

\end{document}